\documentclass[twocolumn,pra,superscriptaddress]{revtex4}
\usepackage{amsmath,amssymb,}
\usepackage{mathrsfs}
\usepackage{graphicx}% Include figure files
\usepackage{dcolumn}% Align table columns on decimal point
\usepackage{bm}% bold math
\usepackage[colorlinks,citecolor=red]{hyperref}

\usepackage{rotating}
\usepackage{floatrow}

\usepackage{lipsum}

\usepackage{bbm}

\begin{document}

\title{Quantum quenches in a pseudo-Hermitian Chern insulator}

\author{Peng He}
\affiliation{Guangdong Provincial Key Laboratory of Quantum Engineering and Quantum Materials, School of Physics and Telecommunication Engineering, South China Normal University, Guangzhou 510006, China}

\affiliation{Department of Physics, Centre for Quantum Coherence, and The Hong Kong Institute of Quantum Information Science and Technology, The Chinese University of Hong Kong, Shatin, New Territories, Hong Kong, China}
%\affiliation{National Laboratory of Solid State Microstructures, Collaborative Innovation Center of Advanced
%Microstructures, Nanjing University, Nanjing 210093, China}

\author{Yan-Qing Zhu}
\affiliation{Guangdong-Hong Kong Joint Laboratory of Quantum Matter, Department of Physics,
and HKU-UCAS Joint Institute for Theoretical and Computational Physics at Hong Kong,
The University of Hong Kong, Pokfulam Road, Hong Kong, China}

\author{Jian-Te Wang}
\affiliation{School of Physics, Nanjing University, Nanjing 210093, China}
%\affiliation{National Laboratory of Solid State Microstructures, Collaborative Innovation Center of Advanced
%Microstructures, Nanjing University, Nanjing 210093, China}

\author{Shi-Liang Zhu}
	\email{slzhu@nju.edu.cn}
\affiliation{Guangdong Provincial Key Laboratory of Quantum Engineering and Quantum Materials, School of Physics and Telecommunication Engineering, South China Normal University, Guangzhou 510006, China}

\affiliation{Guangdong-Hong Kong Joint Laboratory of Quantum Matter, Frontier Research Institute for Physics, South China Normal University, Guangzhou 510006,
China}

\date{\today}

\begin{abstract}

We propose to uncover the topology of a pseudo-Hermitian Chern insulator by quantum quench dynamics. The Bloch Hamiltonian of the pseudo-Hermitian Chern insulator is defined in the basis of the q-deformed Pauli matrices, which are related to the representation of the deformed algebras. We show the bulk-surface duality of the pseudo-Hermitian phases, then further build a concrete relation between the static band topology and quench dynamics, in terms of the time-averaged spin textures. The results are also generalized into a fully nonequilibrium case where the post quench evolution is governed by a Floquet pseudo-Hermitian Hamiltonian. Furthermore, we propose a possible scheme to realize the seemingly challenging model in a bilayer lattice and detect the dynamics with a double-quench protocol.

\end{abstract}
%\pacs{42.50.Pq, 37.30.+i, 03.67.Bg, 76.30.Mi}

\maketitle

\section{Introduction}

Past years have seen a great research interest in the topological theory of open and nonequilibrium systems both theoretically and experimentally \cite{DWZhang2018,Ashida2020,Bergholtz2021}. A quantum system can be driven out of equilibrium either by dynamical control of the parameters or by coupling the system to the environment. Typical ways include Floquet engineering with a periodic driving \cite{Goldman2014,Eckardt2017,Rudner2020}, and quantum quenches following a sudden change of the Hamiltonian \cite{Caio2015,CWang2017,CYang2018,HHu2020,XChen2020,LZhang2018,TLi2021,Jangjan2022}. With the inclusion of the temporal dimension, the topologies of these systems are identified in the momentum-time continuum, leading to unique features transcending equilibrium systems, such as anomalous Floquet insulators whose topology is characterized by the micromotion throughout one evolution period \cite{Kitagawa2010,Rudner2013,Wintersperger2020}, the spacetime Hopf link emergent in a quenched Chern insulator \cite{CWang2017,Unal2019,Ezawa2018,JYu2019}, and the dynamical bulk-surface duality \cite{LZhang2018,LZhang2019,LZhang2020,WJia2021,LZhang2022,ZLei2022}, etc. These findings build a deep connection between the static band topology and quench dynamics, also opening up new routes to characterize topological phases in experimental nonequilibrium settings \cite{WSun2018,Flaschner2018,KWang2019,XTan2019,Unal2019h,HTDing2020,WJi2020}.

Most of the classification schemes with nonequilibrium settings are based on Hermitian Hamiltonians. Recent progress shows that the absence of Hermiticity ramifies new symmetries and predicts exotic topological phases \cite{ZGong2018,Kawabata2019,Leykam2017,HShen20018,SYao2018,KZhang2020,DWZhang2020,QLin2022,LFZhang2021,LZTang2021,Okuma2020,Yokomizo2019}. However, a non-Hermitian Hamiltonian generally has complex spectrum. Modes with a large imaginary part of the eigenenergy will decay fast in time \cite{HWang2018,WZhang2021,YWu2019}. On the other hand, the band topology is defined in terms of the ground state. However, the ground state of a non-Hermitian system could be unstable \cite{Yamamoto2019,YGLiu2019}. Hence dynamical classification not requiring a nontrivial initial state provides a proper way to uncover the topology of the non-Hermitian systems. It is natural to ask if the non-Hermitian systems support similar patterns in quench dynamics as their Hermitian counterparts.

In this paper, we study the quench dynamics in a special class of non-Hermitian systems, namely, the pseudo-Hermitian phases. A pseudo-Hermitian Hamiltonian can be connected to a Hermitian one with a similarity
transformation $H=\eta H^{\dagger} \eta^{-1}$, where $\eta$ is a Hermitian matrix, and thus has a real spectrum and do not support any skin effect \cite{Mostafazadeh2002a,Mostafazadeh2002b,KZhang2020,Okuma2020}. Following Ref. \cite{YQZhu2021}, we construct a pseudo-Hermitian Chern insulator by employing q-deformed Clifford algebra, which provides a general approach to build pseudo-Hermitian phases from the standard Hermitian ones. One can easily obtain a pseudo-Hermitian model by replacing the generators to be the q-deformed ones. The pseudo-Hermitian Chern insulator features a topological quasi-flat band, and is expected to have anomalous transport predicted by the semiclassical Bloch equations \cite{JHWang2022}. We show that the pseudo-Hermitian Chern insulator supports the $\mathbb{Z} \oplus \mathbb{Z}$ bulk-surface duality. On the band inversion surface (BIS), the bulk invariant reduces to two equivalent winding numbers. We develop the nonequilibrium characterization of the bulk-surface duality based on the quench dynamics of the spin textures. The topological index can be extracted by the winding of the time-averaged spin textures across the BIS. Compared to their Hermitian counterparts, the pseudo-Hermitian phases only present dynamical topological patterns on the first-order BIS, but do not have a higher-order characterization. Furthermore, we generalize the model into the Floquet case by periodically switching the mass term, giving a synergy of the three different concepts. The quench of the Floquet pseudo-Hermitian phases gives rise to richer features, identified on both static and driving-induced BISs. We also propose to realize our model with a bilayer lattice system and detect the evolution by a double-quench protocol.

The rest of this paper is organized as follows. In Sec. \ref{sec2}, we define the pseudo-Hermitian Chern insulators and study its topological properties in terms of the BIS. In Sec. \ref{sec3}, we give a dynamical description of the bulk-surface duality with the quench dynamics of the spin textures. In Sec. \ref{sec4}, we study the quench dynamics gorvened by a Floquet pseudo-Hermitian post Hamiltonian. In Sec. \ref{sec5},  we propose a possible lattice realization and detection of the pseudo-Hermitian model. Finally, a short summary is given in Sec. \ref{sec6}.

\section{Topological  characterization}\label{sec2}
We start by considering a two-band Chern insulator whose Hamiltonian is written in the elementary representation matrices of the q-deformed Clifford algebra,
\begin{equation}
\tilde{\mathcal{H}}_{\mathrm{CI}}(\boldsymbol{k})=\mathbf{d}(\mathbf{k})\cdot \boldsymbol{\tilde{\sigma}}=d_{x} \tilde{\sigma}_{x}+d_{y} \tilde{\sigma}_{y}+d_{z} \tilde{\sigma}_{z},\label{eq_ham}
\end{equation}
where the three-component Bloch vector is given as
\begin{equation}
d_{\mu}=v_{\mu} \sin k_{\mu}, d_{z}=v_{z} (M-\sum_{\mu} \cos k_{\mu} ),
\end{equation}
with $\mu=x,y$, the Fermi velocity $v_{x,y,z}=J$,  and $M$ being a tunable parameter with the dimension of energy. Here $\tilde{\sigma}_i$  are the q-deformed Pauli matrices which satisfy the q-deformed Clifford algebra as extended pseudo-Hermitian background symmetry and take the form \cite{Blohmann2003},
\begin{equation}
\tilde{\sigma}_{x}=\left(\begin{array}{ll}
0 & a \\
b & 0
\end{array}\right), \tilde{\sigma}_{y}=\left(\begin{array}{cc}
0 & -i a \\
i b & 0
\end{array}\right), \tilde{\sigma}_{z}=\left(\begin{array}{cc}
q^{-1} & 0 \\
0 & -q
\end{array}\right),
\end{equation}
where $a=\sqrt{(1+q^2)/2}$ and $b=\sqrt{(1+q^{-2})/2}$. $\tilde{\sigma}_i$ form a representation of $\rm SU_q(2)$ algebra \cite{Fujikawa1998},
\begin{equation}
[\tilde{\sigma}_{\pm},\tilde{\sigma_z}]=\mp2\mathbb{I}_q\tilde{\sigma}_{\pm},\quad [\tilde{\sigma}_{+},\tilde{\sigma}_{-}]=\mathbb{I}_q \sigma_z\,,
\end{equation}
and satisfy the q-deformed Clifford algebra,
\begin{equation}
\tilde{\sigma}_{\pm}\tilde{\sigma}_{\pm}=0, \quad \{\tilde{\sigma}_{+},\tilde{\sigma}_{-}\}=\mathbb{I}_q\,,
\end{equation}
where $\tilde{\sigma}_{\pm}=(\tilde{\sigma}_{x}\pm i\tilde{\sigma}_{y})/2$, $\mathbb{I}_q=\frac{[2]}{2}\mathbb{I}$ with $\mathbb{I}$ being the $2\times 2$ identity matrix, and $[x]=(q^x-q^{-x})/(q-q^{-1})$ is the q-number \cite{Bonatsos1999}.

\begin{figure}[htbp]
	\centering
	\includegraphics[width=\textwidth]{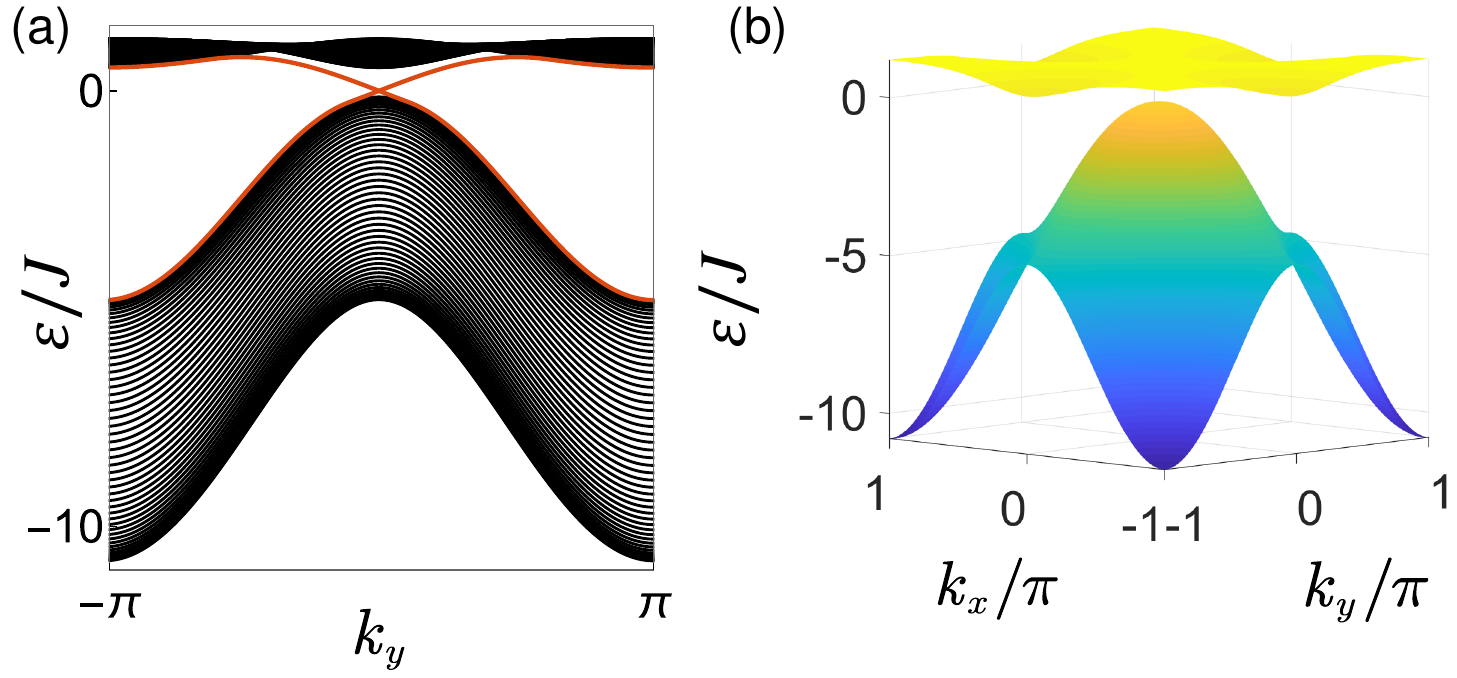}
	\caption{(a) Energy spectrum of the reduced tight-binding chain of the pseudo-Hermitian Chern insulator for $M=0.8J$, $q=3$, and lattice sites $L_x = 50$  when the open boundary condition is imposed along the $z$ direction. (b) The corresponding bulk spectrum of the pseudo-Hermitian Chern insulator.}
	\label{fig_spectrum}
\end{figure}

When $q\neq 1$, this model is pseudo-Hermitian,
\begin{equation}
\eta_{1} \tilde{\mathcal{H}}_{\mathrm{CI}}^{\dagger} \eta_{1}^{-1}=\tilde{\mathcal{H}}_{\mathrm{CI}}, \quad \eta_{1}=\operatorname{diag} (q^{\frac{1}{2}}, q^{-\frac{1}{2}} )\,,
\end{equation}
where $\eta_1^\dagger=\eta_1$ and $(\eta_1^{-1})^\dagger=\eta_1^{-1}$. The Hamiltonian (\ref{eq_ham}) has a real energy spectrum due to the pseudo-Hermiticity as illustrated in Fig. \ref{fig_spectrum}:
\begin{equation}
\varepsilon_{\pm}=d d_{z} \pm \varepsilon=d d_{z} \pm\sqrt{a b\left(d_{x}^{2}+d_{y}^{2}\right)+c^{2} d_{z}^{2}}\,,
\end{equation}
where $c=(1+q^2)/(2q)$ and $d=(1-q^2)/(2q)$. The model in Eq. (\ref{eq_ham}) corresponds to symmetry class A and the topology is characterized by the first Chern number defined through the biorthogonal basis,
\begin{equation}
C_{n}=\frac{1}{2 \pi} \int_{\mathbb{T}^{2}} d k_{\mu} d k_{v} F_{\mu v}^{n}\,,
\end{equation}
where the associated Berry curvature for the $n$th Bloch band is given by
\begin{equation}
F_{\mu \nu}^{n}=\partial_{\mu} A_{v}^{n}-\partial_{v} A_{\mu}^{n}, \quad A_{\mu}^{n}=\langle u_{n}^{L}|i \partial_{\mu}| u_{n}^{R}\rangle\,,\label{eq_curc}
\end{equation}
with $\langle u^L_{n}|$ and $| u^R_{n}\rangle$ being the $n$th left and right eigenvectors, respectively. For the two-band model considered here, the Chern number is equivalent to the integral of the (generalized) Bloch spin vector field (for more details, see Appendix \ref{appa}):
\begin{equation}
C_n=\frac{1}{4 \pi} \int_{\mathbb{T}^{2}} d k_{\mu} d k_{v}~ \epsilon_{ijk} h_{i} \partial_{\mu} h_{j} \partial_{\nu} h_{k}/\varepsilon^3,\label{eq_cn}
\end{equation}
where $h_x=\frac{a+b}{2}d_x-i\frac{a-b}{2}d_y$, $h_y=i\frac{a-b}{2}d_x+\frac{a+b}{2}d_y$, and $h_z=cd_z$ are the components of the Bloch vector decomposed in terms of the Pauli matrices $\sigma_{x,y,z}$, and $h=\sqrt{h_x^2+h_y^2+h_z^2}$. For $|M| < 2$ ($|M| > 2$),
$C_1 = -{\rm sgn}(M)~(C_1 = 0)$ for the lower band.

To facilitate the description, we reduce the topological characterization by the Chern number to a one dimensional winding number, adopting the method based on the BIS. Following the conventions used in Ref. \cite{LZhang2018}, we take the even-parity component $d_z$ to define the dispersion of the decoupled bands, and denote the remaining components as the spin-orbit (SO) field $\boldsymbol{d}_{\rm{so}}(\boldsymbol{k})=(d_x,d_y)$. The BIS is defined as all band crossings for the decoupled bands, i.e., $d_z(\boldsymbol{k})=0$, forming a closed one-dimensional surface. The Chern number of the bulk band reduces to a  one-dimensional winding number on the BIS \cite{LZhang2018,Ghatak2019}:
\begin{equation}
C_1(\boldsymbol{d}_{\rm{so}},d_z)|_{\mathbb{T}^{2}}=w_1(\boldsymbol{d}_{\rm{so}})|_{\rm{BIS}}\,.
\end{equation}
We show that the SO field $\boldsymbol{d}_{\rm{so}}(\boldsymbol{k})$ gives out the same winding number as the deformed SO field $\tilde{\boldsymbol{d}}_{\rm{so}}(\boldsymbol{k})$:
\begin{equation}
w_1(\boldsymbol{d}_{\rm{so}})=\frac{1}{2\pi}\int_{\rm{BIS}} \hat{\boldsymbol{h}}_{\rm{so}} d \hat{\boldsymbol{h}}_{\rm{so}}=\frac{1}{2\pi}\int_{\rm{BIS}} \hat{\boldsymbol{d}}_{\rm{so}} d \hat{\boldsymbol{d}}_{\rm{so}}\,,\label{eq_wn}
\end{equation}
where $\hat{\boldsymbol{h}}_{\rm{so}}=\boldsymbol{h}_{\rm{so}}/|\boldsymbol{h}_{\rm{so}}|$ and $\hat{\boldsymbol{d}}_{\rm{so}}=\boldsymbol{d}_{\rm{so}}/|\boldsymbol{d}_{\rm{so}}|$ (for more details of the proof, see Appendix \ref{appb}). From above discussions, we see that the pseudo-Hermitian phases have corresponding equivalent Hermitian copies, thus supporting the $\mathbb{Z} \oplus \mathbb{Z}$ bulk-surface duality.

\section{Quench dynamics}\label{sec3}
\begin{figure}[htbp]
	\centering
	\includegraphics[width=\textwidth]{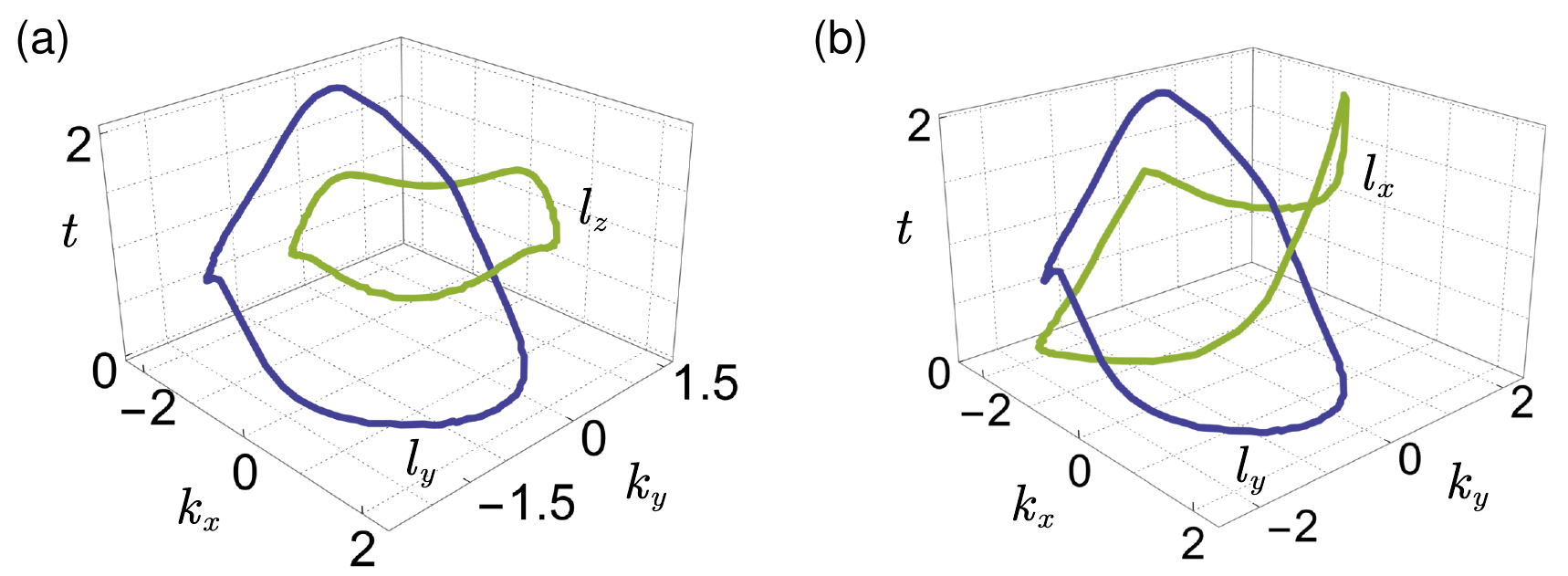}
	\caption{The Hopf link formed by the preimages of two time-evolved states in momentum-time space. Here $l_x=(1,0,0)$, $l_y=(0,1,0)$, and $l_z=(0,0,1)$.}
	\label{fig_link}
\end{figure}

We proceed to consider characterization of the pseudo-Hermitian Chern insulator by a sudden quench of the mass term in the Hamiltonian (\ref{eq_ham}) from a deep initially trivial phase at time $t = 0$ to a topological phase. For the pre-quench Hamiltonian $\mathcal{H}_i$, $M|_{t<0}\gg\max[|\mathbf{d}(\mathbf{k})|]$, whereas $M|_{t>0}<\max[|\mathbf{d}(\mathbf{k})|]$ for the post-quench Hamiltonian $\mathcal{H}_f$.  The system is initialized in the ground state $|\xi_i(\boldsymbol{k})\rangle$ of the pre-quench Hamiltonian. After the sudden quench, the state evolves into $|\xi(\boldsymbol{k},t)\rangle=e^{-i\mathcal{H}_f(\boldsymbol{k})t}|\xi_i(\boldsymbol{k})\rangle$ at some later time $t$. The quench dynamics has period $\pi/\varepsilon$ since $|\xi(\boldsymbol{k},\pi/\varepsilon)\rangle=-e^{ic d_z\pi/\varepsilon}|\xi_i(\boldsymbol{k})\rangle$  returns to the initial state up to an overall phase. For the pseudo-Hermitian system, we can also construct a Bloch vector according to $\boldsymbol{n}=\langle\xi| \boldsymbol{\sigma} |\xi \rangle/\langle\xi|\xi\rangle$, which defines a map from $\mathbb{T}^{3}$ to $\mathbb{S}^2$. Then the quench dynamics can be  characterized by the Hopf invariant \cite{CWang2017}:
\begin{equation}
\mathcal{L}=\int_{\mathbb{T}^{3}} d^2\boldsymbol{k}dt \tilde{\boldsymbol{A}}\cdot \tilde{\boldsymbol{F}}\,,
\end{equation}
where $\tilde{F}_\mu=\frac{1}{8\pi}\epsilon^{\mu\nu\rho}\boldsymbol{n}\cdot(\partial_\nu \boldsymbol{n}\times \partial_\rho \boldsymbol{n})$ and $\nabla \times \tilde{\boldsymbol{A}}=\tilde{\boldsymbol{F}}$. For a nontrivial Hopf map, the preimage of a point given by $\boldsymbol{n}$ in $\mathbb{S}^2$ forms a closed loop in $\mathbb{T}^{3}$, and the linking number of two such loops corresponds to the Hopf invariant, providing a way to detect the Hopf invariant. In Fig. \ref{fig_link}, we show two typical cases. For the non-trivial phase with $\mathcal{L}=1$, the preimages of two arbitrary states form a Hopf link. The linking structure can be measured in cold-atom experiments with time-of-flight imaging (TOF), as we see in the following discussions.

We further consider the characterization of the bulk-surface duality of the pseudo-Hermitian phase. Specifically, we focus on the long-time average of the Bloch vector $\boldsymbol{n}$:
\begin{equation}
\bar{\boldsymbol{n}}=\lim_{T\to\infty}\frac{1}{T} \int_{0}^{T}dt~\langle\xi| \boldsymbol{\sigma} |\xi \rangle/\langle\xi|\xi\rangle\,.\label{eq_texture}
\end{equation}
Pictorially, the post-quench evolution is analogous to the procession of the spin $\xi$ with respect to the Bloch vector field $\mathcal{H}_f$. The initial configuration of $\xi(\boldsymbol{k})$ corresponds to a fully polarized spin assigned on every generic momentum along the direction of $d_z$. On the BIS, $\xi(\boldsymbol{k})_0$ is orthogonal to the Bloch field $\boldsymbol{d}_{\rm{so}}$, giving rise to vanishing time-averaged Bloch vector $\bar{\boldsymbol{n}}$ (for the rigorous proof based on a generic two-band model, see the Appendix \ref{appc}). Hence the BIS can be determined by the dynamics of $\bar{\boldsymbol{n}}$:
\begin{equation}
\bar{\boldsymbol{n}}_i=0\,,~{\rm{for}}~\boldsymbol{k}\in {\rm BIS}\,, ~i=x, y.
\end{equation}
Furthermore, the topological index given by Eq. (\ref{eq_wn}) is extracted by the variation of the time-averaged Bloch vector across the BIS. To capture the features across the BIS, we use a dynamical spin-texture field $\boldsymbol{g}(\boldsymbol{k})=-\partial_{k_{\perp}}\bar{\boldsymbol{n}}(\boldsymbol{k})$, where ${k_{\perp}}$ is the momentum perpendicular to the BIS \cite{LZhang2018}. By substituting Eq. (\ref{eq_texture}), we show that on the BIS (for more details, see Appendix \ref{appc})
\begin{equation}
\begin{split}
g_{x,y}(\boldsymbol{k})|_{\rm{BIS}}={\rm Re}(\hat{h}_{x,y})\pm{\rm Im}(\hat{h}_{y,x})=\hat{d}_{x,y}\,,
\end{split}
\end{equation}
thus the relation between the topological index defined on the BIS and the winding number of the dynamical spin-texture field holds for both Hermitian and pseudo-Hermitian phases:
\begin{equation}
w_1(\boldsymbol{g})=w_1(\boldsymbol{d}_{\rm{so}}).
\end{equation}

\begin{figure}[htbp]
	\centering
	\includegraphics[width=\textwidth]{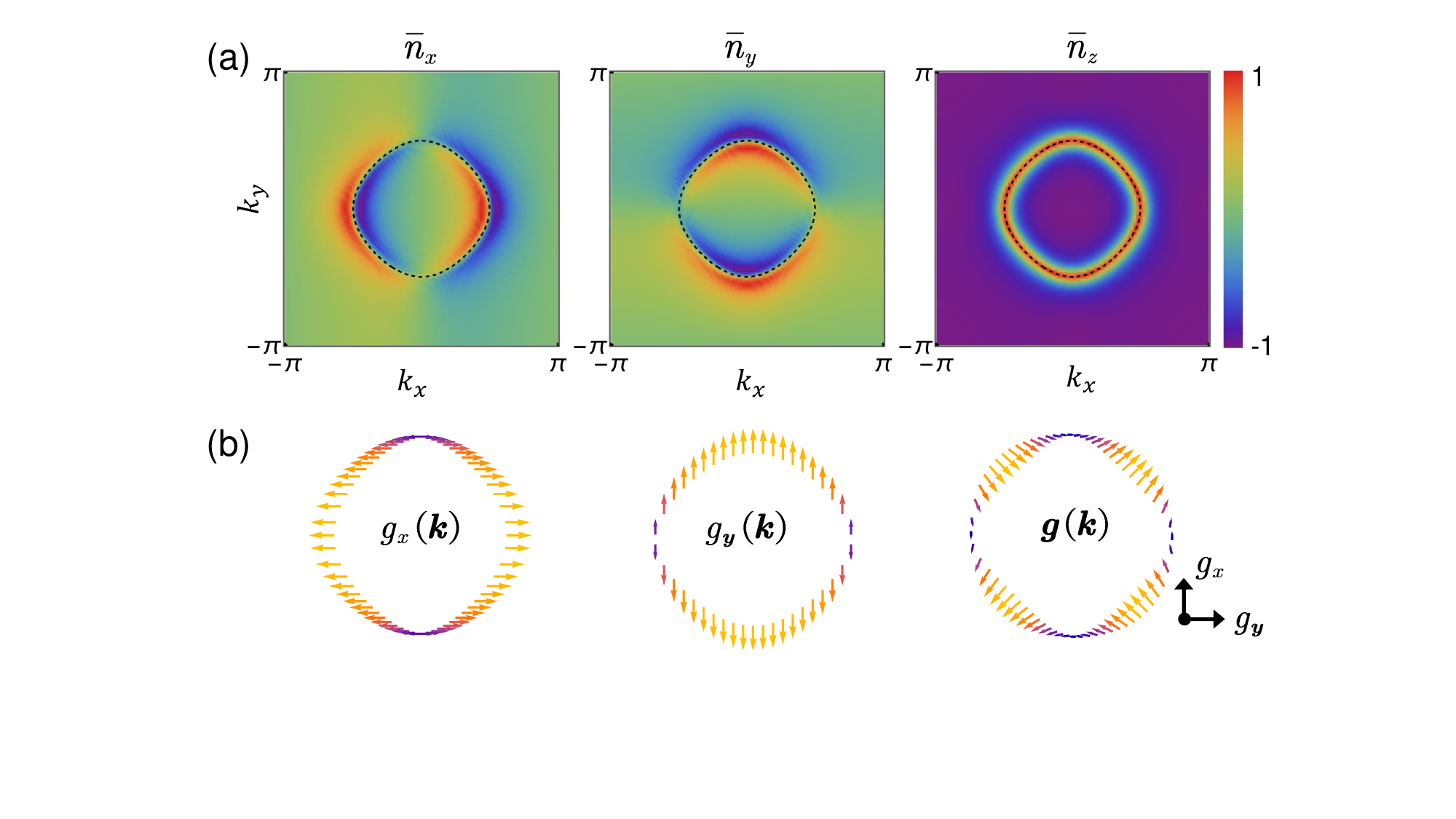}
	\caption{(a) The  long-time average of the Bloch vector $\boldsymbol{n}$ after a sudden quench from $M=\infty$ to $M=J$. The dashed lines indicate the BIS determined by $d_z(\mathbf{k})=0$. (b) The dynamical spin-texture field $\boldsymbol{g}(\boldsymbol{k})$ on the BIS. The winding of $\boldsymbol{g}(\boldsymbol{k})$ indicates a nontrivial Chern number $C_1=-1$. The evolution time is taken as $T=30$  (in units of $1/J$).}
	\label{fig_texture}
\end{figure}

In Fig. \ref{fig_texture}, we numerically calculate the averaged Bloch vector $\bar{\boldsymbol{n}}$. The BIS can be determined by the momentum points where the components of the Bloch vector $n_x$ ($n_y$) revise the sign. Fig. \ref{fig_texture}(b) illustrates the winding of the dynamical spin-texture field $\boldsymbol{g}(\boldsymbol{k})$ on the BIS. The field $\boldsymbol{g}(\boldsymbol{k})$ winds once, characterizing a winding number $w_1=-1$, which is consistent with the theoritical prediction.

\section{Floquet Pseudo-Hermitian Phases}\label{sec4}

\begin{figure}[htbp]
	\centering
	\includegraphics[width=\textwidth]{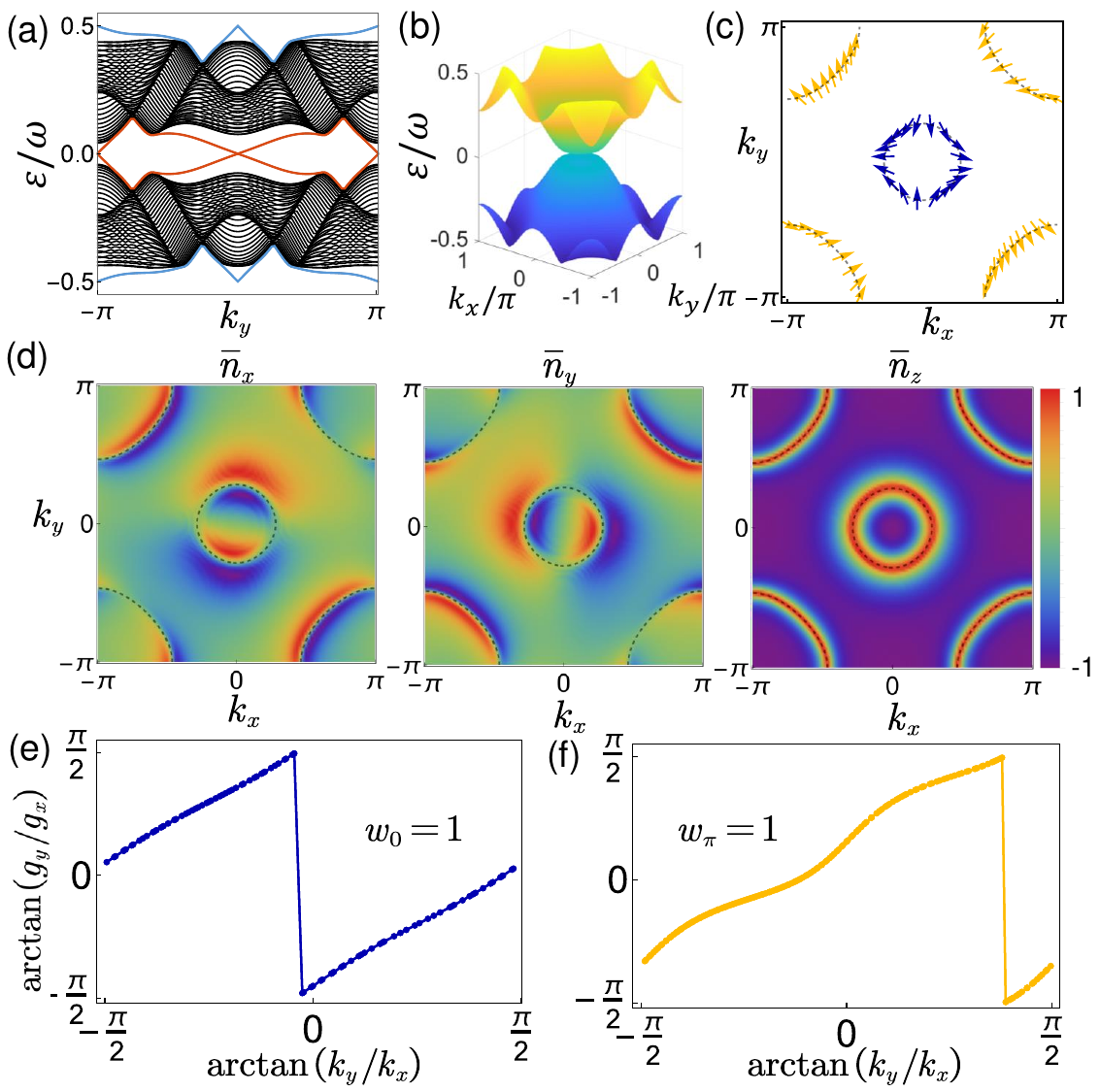}
	\caption{(a) Quasi-energy spectrum of the Floquet pseudo-Hermitian Chern insulator with open boundary condition imposed along the $z$ direction. (b) The band structure of the post-quench effective Floquet Hamiltonian  $\mathcal{H}_F$. (c) The winding of the dynamical field $\mathbf{g}(\mathbf{k})$ on the 0 BIS and $\pi$ BIS. (d) The the long-time average of the Bloch vector $\boldsymbol{n}$ after a sudden quench from a deeply trivial phase to a nontrivial Floquet pseudo-Hermitian phase. (e) and (f) The relative phase angle of the dynamical field $g_x(\mathbf{k})$ and $g_y(\mathbf{k})$ on the 0 BIS and $\pi$ BIS, which characterizes the topological invariants $w_{0,\pi}=1$. The parameters in the post-quench Hamiltonian are as follows: $M=8v_z$, $q=3$, $\omega=5v_z$, $T_1=0.6T$, $v_{x,y,z}=J$. }
	\label{fig_floquet}
\end{figure}

In this section, we extend the results to the Floquet case. Specifically, we focus on the polychromatic driving of the mass term in the Hamiltonian (\ref{eq_ham}) $V(t)=M(t)\sigma_z$, where
\begin{equation}
M(t)= \left\{\begin{array}{lc}
M & 0 \leq t<T_{1} \\
-M & T_{1} \leq t<T\,.
\end{array}\right.\label{eq_drive}
\end{equation}
Here for simplicity we only take the representation matrices of the spin-orbit coupling term to be the q-deformed matrices. The quasi-energy band $\varepsilon$ of the driven system is characterized by an effective Hamiltonian $\mathcal{H}_F=i\log U(T)/T$, where the evolution operator $U(T)\equiv \mathcal{T} \exp  [-i \int_{0}^{t} H(t) d t ]$ and $\mathcal{T}$ denotes the time ordering. We take the quasienergy $\varepsilon \in[-\pi / T, \pi / T]$ as the first Floquet Brillouin zone (FBZ). As a result, the in-gap modes can be found at either $\varepsilon=0$ or $\varepsilon=\pi/T$, as shown in Fig. \ref{fig_floquet}(a) for the quasienergy spectrum in the FBZ with a cylindrical geometry. Now we consider the quench dynamics for the post-quench Floquet pseudo-Hermitian Hamiltonian. The quasienergy $\varepsilon$ is only defined up to integer multiples of the driving energy quantum $m\omega$, thus the Floquet Hamiltonian enables two types of the BIS according to the odevity of $\delta m\equiv |m_1-m_2|$, i.e., $\pi$ BIS for odd $\delta m$ while 0 BIS for even $\delta m$, where $m_1$ and $m_2$ are the indices of the two subbands. Then the lower Floquet band is characterized by the winding number:
\begin{equation}
W=\sum_{j}(-1)^{q_{j} / \pi} w_{j}\,,
\end{equation}
where $q_j=0$ or $\pi$ for the $q_j$ BIS.

For the Floquet system, the dynamical  bulk-surface duality is revealed by the stroboscopic time-averaged Bloch vector:
\begin{equation}
\overline{  n}_{i}(\boldsymbol{k}) =\lim _{N \rightarrow \infty} \frac{1}{N} \sum_{n=0}^{N}  n_{i}(\boldsymbol{k}, t=n T) , \quad i=x, y, z.
\end{equation}
Here $n_i(\boldsymbol{k}, t)=\langle\xi(\boldsymbol{k}, t)| \boldsymbol{\sigma} |\xi(\boldsymbol{k}, t) \rangle/\langle\xi(\boldsymbol{k}, t)|\xi(\boldsymbol{k}, t)\rangle$. We show the numerical results for the Floquet pseudo-Hermitian phase in Fig. \ref{fig_floquet}(c)-(f). The averaged Bloch vector $\overline{  n}_{x,y}$ vanishes at two rings centered at the center and corner of the first Brillouin zone, identified as 0 BIS and $\pi$ BIS respectively (the analytic solution of the BIS is given in Appendix \ref{appd}). The two BISs distinguish more clearly in terms of the band structure, as shown in Fig. \ref{fig_floquet}(b). The winding number $W$ is characterized by the dynamical field extracted from the stroboscopic time-averaged Bloch vector across the BIS. $\mathbf{g}(\mathbf{k})$ winds once both on the 0 BIS and $\pi$ BIS, implying the invariants $w_{0,\pi}=+1$ [the sign is readily obtained from Fig. \ref{fig_floquet}(e) and (f)]. The Chern number is $C_1=w_0-w_{\pi}=0$, which means the system is in an  anomalous Floquet pseudo-Hermitian phase.

\section{Lattice Realization and Detection}\label{sec5}

\begin{figure}[htbp]
	\centering
	\includegraphics[width=\textwidth]{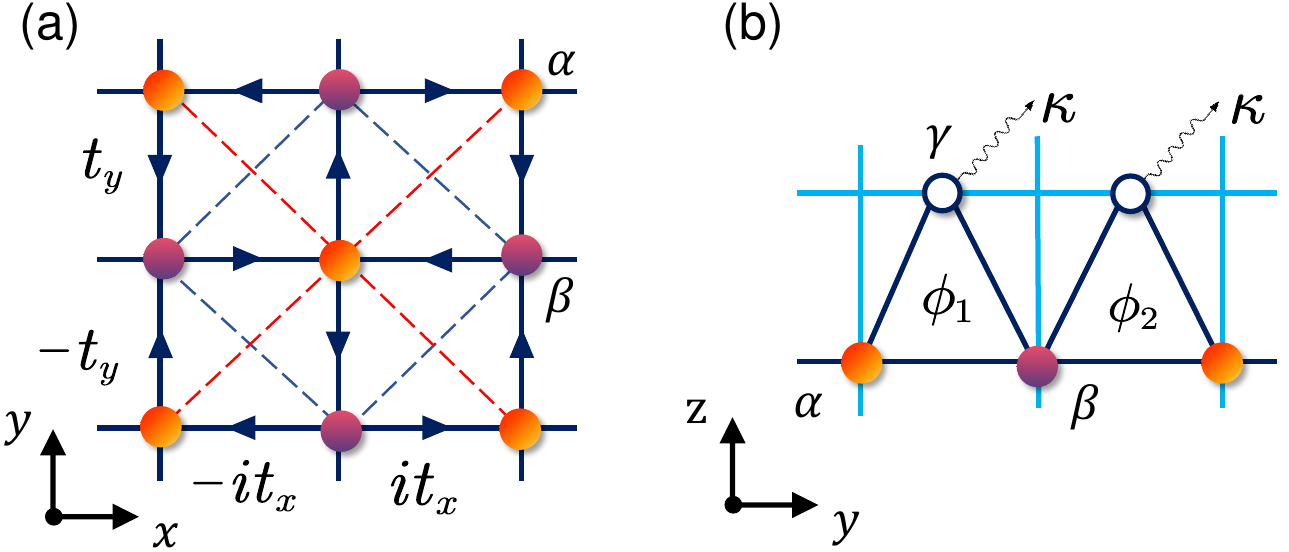}
	\caption{Schematic illustration of the bilayer cubic lattice with engineered hopping on (a) the $x$-$y$ plane and (b) the $y$-$z$ plane. The dashed lines in (a) show the next-nearest-neighbor hoppings. Sites $\gamma$ in the upper layer are dissipative, used as auxiliary sites to induce asymmetric coupling. The phases of the inter-layer hopping rates contribute a staggered synthetic flux $\phi_1$ and $\phi_2$.}
	\label{fig_sch}
\end{figure}

The realization of the Hamiltonian (\ref{eq_ham}) on a lattice requires asymmetric coupling and is thus rather challenging in a realistic experiment. In this section we propose one way to demonstrate the pseudo-Hermitian Chern insulator with ultracold atoms in a bilayer optical lattice. A primary lattice is coupled with a dissipative auxiliary lattice, as illustrated in  Fig. \ref{fig_sch}. The full tight-binding Hamiltonian is given by
\begin{equation}
\mathcal{H}=\mathcal{H}_{p}+\mathcal{H}_{a}+\mathcal{H}_{p-a}\,,\label{eq_latt}
\end{equation}
\begin{equation}
\begin{aligned}
\mathcal{H}_p=&\sum_{\langle\vec{i}, \vec{j}\rangle}\sum_{ \mu,\mu'=\alpha, \beta}  J_{i j} (\mathrm{i} \nu_{i j} e^{-i\phi^{(0)}_{ij}} ) c_{\mu, i}^{\dagger} c_{\mu', j} \\
&-\sum_{\langle\langle\vec{i}, \vec{j}\rangle\rangle } \sum_{ \mu=\alpha, \beta} J_{z} c_{\mu, i}^{\dagger} c_{\mu, j}+m_{z} \sum_{i} (n_{i, \alpha}-n_{i, \beta} ),
\end{aligned}
\end{equation}
\begin{equation}
\mathcal{H}_{a}=\sum_{i} (\Delta-i\kappa)n_{i,\gamma},
\end{equation}
\begin{equation}
\begin{split}
\mathcal{H}_{p-a}&=\sum_{i} J_{s} e^{i \phi_1} c_{i,\gamma}^{\dagger} c_{i,\alpha}+J_{r} c_{i,\gamma}^{\dagger} c_{i+1_y,\beta}\\
&+J_{s} e^{i \phi_1} c_{i,\gamma}^{\dagger} c_{i,\beta}+J_{r} c_{i,\gamma}^{\dagger} c_{i+1_y,\alpha}+\rm{H.c.},
\end{split}
\end{equation}
where $n_{i,\mu}=c_{i,\mu}^{\dagger} c_{i,\mu}$, with $c_{i,\mu}^{\dagger}$ and $c_{i,\mu}$ the creation and annihilation operator, the notations $\langle\vec{i}, \vec{j}\rangle$ and $\langle\langle\vec{i}, \vec{j}\rangle\rangle$ running over all nearest- and next-nearest-neighbor sites, $\nu_{i j}=1(-1)$ for
hopping along (opposite to) the marked direction in Fig. \ref{fig_sch}(a), and $\phi^{(0)}_{ij}=\pi (0)$ for $j=i+1_x(j=1+1_y)$. When $J_{s}, J_{r} \ll|\Delta-i \gamma|$, the Hamiltonian (\ref{eq_latt}) can be
reduced to a non-Hermitian one with asymmetric coupling by projection into the subspace $\{|i,\alpha\rangle,|i,\beta\rangle\}$, and the non-Hermitian terms read \cite{TLiu2019,QLiang2022}
\begin{widetext}
\begin{equation}
\begin{split}
\mathcal{H}_{\rm{NH}}&=\sum_i (-J_x-e^{-i\phi_1}\tilde{\kappa})c_{\alpha, i}^{\dagger} c_{\beta, i+1_x}+(J_y-e^{-i\phi_2}\tilde{\kappa})c_{\alpha, i}^{\dagger} c_{\beta, i-1_y}+(-J_y-e^{i\phi_1}\tilde{\kappa})c_{\beta, i+1_x}^{\dagger} c_{\alpha, i}\\
&+(J_y-e^{i\phi_2}\tilde{\kappa})c_{\beta, i-1_y}^{\dagger} c_{\alpha, i}-\frac{J_{s}^{2}(\Delta+i \gamma)}{\Delta^{2}+\gamma^{2}} n_{i,\alpha}-\frac{J_{r}^{2}(\Delta+i \gamma)}{\Delta^{2}+\gamma^{2}} n_{i,\beta},
\end{split}
\end{equation}
\end{widetext}
where $\tilde{\kappa}=\frac{J_{s} J_{r}(\Delta+i \gamma)}{\Delta^{2}+\gamma^{2}}$. In the following derivations, we set $J_s=J_r=\bar{J}$, $\Delta=0$, and $\phi_1=\phi_2+\pi=\pi/2$. After Fourier transformation, the Hamiltonian in $\mathbf{k}$ space can be written as $\hat{H}_{\mathbf{k}}=\sum_{\mathbf{k}, \mu \mu^{\prime}} c_{\mathbf{k} \mu}^{\dagger}[\mathcal{H}(\mathbf{k})]_{\mu \mu^{\prime}} \boldsymbol{c}_{\mathbf{k} \mu^{\prime}}$, where $c_{\boldsymbol{k} \mu}=1 / \sqrt{V} \sum_{\boldsymbol{r}} e^{-i \boldsymbol{k} \cdot \boldsymbol{r}} c_{\boldsymbol{r} \mu}$ is the annihilation operator in $\mathbf{k}$ space, and the Bloch Hamiltonian reads,
\begin{equation}
\begin{split}
\mathcal{H}(\mathbf{k})=&\bar{\kappa}\sigma_0+J_x/2\sin k_x\sigma_x-\bar{J}_y/2\sin k_y\tilde{\sigma}_y\\
&+J_z/2\cos k_x\cos k_y\sigma_z,
\end{split}\label{eq_sy}
\end{equation}
with $\bar{\kappa}=\bar{J}^2/\kappa$, $\bar{J}_ya=J_y+\tilde{\kappa}$, and $\bar{J}_yb=J_y-\tilde{\kappa}$. The first term corresponds to a background dissipation. Hamiltonian (\ref{eq_sy}) is a pseudo-Hermitian Chern insulator when we drop the background dissipation term which does not alert the band topology.

The lattice model Eq. (\ref{eq_latt}) can be realized with ultracold atoms in an optical lattice. The involved atomic hoppings can be achieved by the laser-assisted tunneling technique with well-designed Raman
coupling. A feasible scheme to realize the primary lattice has been proposed in a recent work with a  double-well Raman lattice \cite{XJLiu2016}. The bilayer structure can be generated by a set of spin-dependent lattice potential when we use two distinct spin states encoding the primary and auxiliary sites respectively \cite{Tudela2019,ZMeng2021}. The non-Hermitian term can be induced by applying a radio frequency pulse to resonantly transfer the atoms on the auxiliary sites to an irrelevant excited state \cite{JLi2019,Naghiloo2019}. In the experiment, we can control the inter- and intralayer hopping
ratios $J_{s,r}/J_{y}$ with the lattice depth and Raman coupling. Here we assume $J_{s,r}=J_{y}\approx 2\pi\times 100~\rm{Hz}$, and a large dissipation rate $\kappa=5 J_{s,r}$, then we can generate a non-Hermiticity $q=1.5$ in this system \cite{Tudela2019}.

\begin{figure}[htbp]
	\centering
	\includegraphics[width=\textwidth]{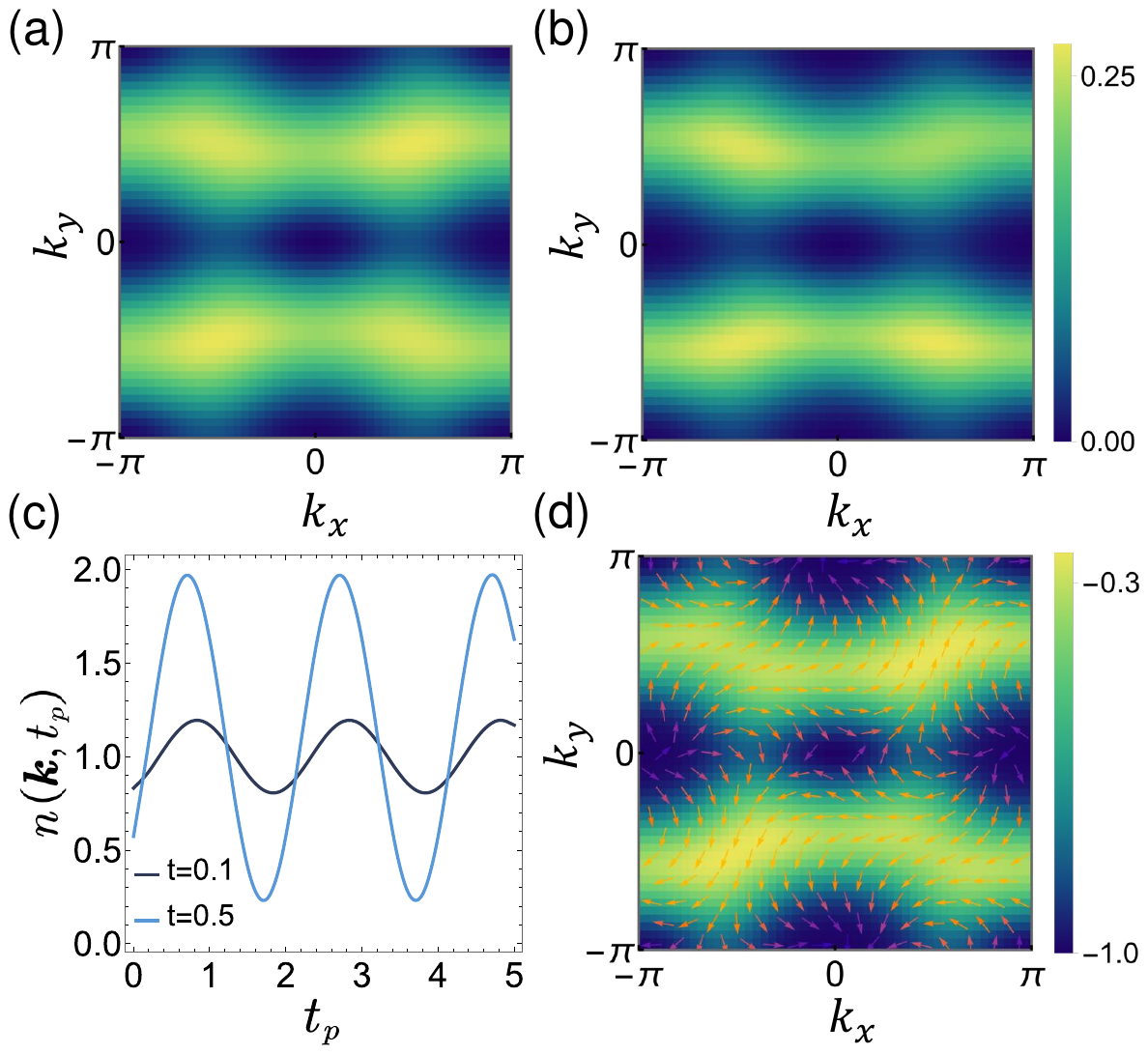}
	\caption{(a) The spectroscopic signatures of the momentum density $n_\alpha(\mathbf{k},t)$ for Hamiltonian (\ref{eq_sy}) after the first quench at time $t=0.5$ (in units of $1/J$). (b) the corresponding tomography results of $n_\alpha(\mathbf{k},t)$ extracted from the TOF images after the second quench, using $30$ time sampling points. (c) Density oscillations with respect to the precession time $t_p$ at momentum point $\mathbf{k}=(\pi/2,\pi/2)$ after the second quench. (d) Example image of all components of the Bloch vector $n_i$ ($i=x,y,z$) in the $k_x$-$k_y$ plane. The arrows show the winding of the in-plane components $(n_x,n_y)$, and the background color scale shows the magnitude of the out-of-plane component $n_z$. The other parameters are as follows: $v_z=5v_{x,y}=2J$, $m=J$, $q=3$, $\Delta_{\alpha\beta}=\pi$.   }
	\label{fig_tomo}
\end{figure}

Next, we show that the links and dynamical fields discussed in Sec. \ref{sec3} can be measured in
cold-atom experiments with a time-resolved state tomography, which involves a double quench protocol \cite{Hauke2014,Tarnowski2019}. We assume the tomography Hamiltonian after the second quench is dispersionless, $\mathcal{H}_f=\frac{\Delta_{\alpha\beta}}{2} \sigma_z$, then the ensuing evolution of $\xi$ after TOF expansion translates into the precession on the Bloch sphere. We express the evolved state $\xi(\mathbf{k},t)$ in a general form $\xi(\mathbf{k},t)=(\cos \frac{\theta_{\mathbf{k},t}}{2},\cos \frac{\theta_{\mathbf{k},t}}{2} e^{i\phi_{\mathbf{k},t}})^{\rm{T}}$, then the momentum distribution after a time-of-flight measurement reads,
\begin{equation}
n_p\left(\boldsymbol{k}, t^{p}\right)=f(\boldsymbol{k})\left[1+\sin \theta_{\boldsymbol{k}, t} \cos \left(\Delta_{\alpha\beta} t^{p}+\phi_{\boldsymbol{k}, t}\right)\right],
\end{equation}
where $f(\mathbf{k})$ is the broad envelope function given by the momentum distribution of the Wannier function, and $t^p$ is the precession time.
%The density oscillations as time $t^p$ varies directly reveals both $\phi_{\mathbf{k}}$ and  $\sin \theta_{\mathbf{k}}$.
In Fig. \ref{fig_tomo}(b), we show the numerical simulation of the  momentum density $n_\alpha(\mathbf{k})$ for sublattice $\alpha$, which is reconstructed from $\theta_{\mathbf{k}}$ and $\phi_{\mathbf{k}}$ obtained by TOF images. The tomography result is in good agreement with that calculated with the target state [shown in Fig. \ref{fig_tomo}(a)]. In principle, we can obtain the Bloch vector $n_i$ ($i=x,y,z$) with the state tomography, as shown in Fig. \ref{fig_tomo}(d).

We adopt an $\epsilon$-neighborhood method to find the preimages of a certain Bloch vector orientation \cite{DLDeng2018}. Without loss of generality, we take $l_x=(1,0,0)$ for instance. In a real experiment, $\mathbf{n}$ is always measured with a finite resolution. We need to search $l_x$ in a small $\epsilon$ neighborhood:
\begin{equation}
N_{\epsilon} (l_{x})=\left\{ \mathbf{n}: | \mathbf{n}-l_{x} | \leq \epsilon\right\},\label{eq_ne}
\end{equation}
where $| \mathbf{n}-l_{x} |=[(n_x-l_{x,1})^2+(n_y-l_{x,1})^2+(n_z-l_{x,3})]^{1/2}$ defines the distance between $\mathbf{n}$ and $l_x$. The preimages are the momentum-time points at which the states fulfill the condition Eq. (\ref{eq_ne}). By choosing the proper value of $\epsilon$, we can resolve the ring structure with desired tolerance.

\section{Conclusions}\label{sec6}
In summary, we have investigated the quench dynamics of a pseudo-Hermitian Chern insulator. We show that the pseudo-Hermitian phases support the $\mathbb{Z} \oplus \mathbb{Z}$ bulk-surface duality: the Chern number of the bulk bands is related with two equivalent winding numbers on the BIS. We have presented the nonequilibrium characterization of this duality in the form of the quench dynamics with Hermitian observables, i.e., time-averaged spin textures. We have also extended our discussion to the case of Floquet pseudo-Hermitian systems, which show richer patterns in the spin textures due to the periodic driving. The special construction based on q-deformed Pauli matrices makes the experimental realization rather challenging. We have proposed a possible scheme based on the bilayer optical lattice and practical detection method. The model we adopted has real spectra and thus closes the point gap, leading to the absence of the non-Hermitian skin effect. By breaking the pseudo-Hermiticity, one can study the skin
effect in the previous model. The skin effect will lead to a non-Bloch description \cite{Yokomizo2019}, which is an interesting direction for further exploration. And our models can be generalized to other symmetry classes, to study the quench dynamics of non-Hermitian systems in a myriad of topical issues.

\acknowledgments
This work was supported by the Key-Area Research and Development Program of GuangDong Province (Grant No. 2019B030330001) and the National Natural Science Foundation of China (Grants  No. 12074180 and No. U1801661).

\begin{appendix}
\section{Pauli Matrix Representation}\label{appa}
To calculate the Chern number, we rewrite the Hamiltonian (\ref{eq_ham}) in the standard Pauli matrix basis,
\begin{equation}
\mathcal{H}_{CI}=d d_z\sigma_0+\mathbf{h}(\mathbf{k})\cdot\boldsymbol{\sigma},
\end{equation}
where $\sigma_0$ is a $2\times 2$ identity matrix and $\mathbf{h}=(h_x,h_y,h_z)$, with $h_x=\frac{a+b}{2}d_x-i\frac{a-b}{2}d_y$, $h_y=i\frac{a-b}{2}d_x+\frac{a+b}{2}d_y$, and $h_z=cd_z$. We can drop the first term as it does not alert the topology. The eigenstates are given by
\begin{equation}
|u_{n}^{R}\rangle=(h_x-ih_y,\varepsilon_{n}-h_z)^{\rm{T}}/\sqrt{2\varepsilon_{n}(\varepsilon_{n}-h_z)}\,,
\end{equation}
\begin{equation}
\langle u_{n}^{L}|=(h_x+ih_y,\varepsilon_{n}-h_z)/\sqrt{2\varepsilon_{n}(\varepsilon_{n}-h_z)}\,,
\end{equation}
where $\varepsilon_n=n\sqrt{\mathbf{h} \cdot \mathbf{h}} $, and $n=\pm$ is the band index. Direct calculation of Eq. (\ref{eq_curc}) shows that
\begin{equation}
A_{\mu}^n=\frac{1}{2\varepsilon_{n}(\varepsilon_{n}-\tilde{d_z})}\epsilon_{ij} h_i\partial_\mu h_j,
\end{equation}
\begin{equation}
F^{n}_{\mu\nu}=\epsilon_{ijk} h_{i} \partial_{\mu} h_{j} \partial_{\nu} h_{k}/2\varepsilon_n^3,
\end{equation}
which just gives Eq. (\ref{eq_cn}).

\section{The equivalence of the two winding numbers}\label{appb}
For convenience, we use an equivalent form of the winding number defined in Eq. (\ref{eq_wn}):
\begin{equation}
w_1=\frac{1}{2\pi}\int_{\rm{BIS}} \partial_{\mathbf{k}} \phi_{xy}(\mathbf{k})\cdot d\mathbf{k}\,,
\end{equation}
where $\phi_{xy}(\mathbf{k})=\arctan[\tilde{d}_y(\mathbf{k})/\tilde{d}_x(\mathbf{k})]$. The imaginary part of $\phi_{xy}$ is a real, continuous function and thus does not contribute to the integral of the winding number. We decompose the winding number as
\begin{equation}
\begin{split}
w_1=&\frac{1}{2\pi}\int_{\rm{BIS}} \partial_{\mathbf{k}} {\rm{Re}}[\phi_{xy}(\mathbf{k})]\cdot d\mathbf{k}\\
=& \frac{1}{4\pi}\int_{\rm{BIS}} \partial_{\mathbf{k}}  [\phi_{xy}^{1}(\mathbf{k})+\phi_{xy}^{2}(\mathbf{k})]\cdot d\mathbf{k}\,,
\end{split}
\end{equation}
where we have used the relation $\tan[2{\rm{Re}}(\phi_{xy})]=\tan[\phi_{xy}^1+\phi_{xy}^2]$ \cite{CYin2018,HWang2019},
\begin{equation}
\begin{split}
\tan \phi_{xy}^1=&\frac{\operatorname{Re} (h_{y} )+\operatorname{Im} (h_{x} )}{\operatorname{Re} (h_{x})-\operatorname{Im} (h_{y})}=\frac{d_y}{d_x}\,,\\
\tan \phi_{xy}^2=&\frac{\operatorname{Re} (h_{y} )-\operatorname{Im} (h_{x} )}{\operatorname{Re} (h_{x})+\operatorname{Im} (h_{y})}=\frac{d_y}{d_x}\,,
\end{split}
\end{equation}
thus we have $w_1[\boldsymbol{d}_{\rm{so}}]=w_1[ \boldsymbol{h}_{\rm{so}}]$.

\section{Spin textures}\label{appc}
We decompose the evolved state in terms of the eigenmodes of the post-quench Hamiltonian $\mathcal{H}_f$:
\begin{equation}
|\xi(\boldsymbol{k})\rangle=\sum_{\mu} c_{\mu}(\boldsymbol{k}) e^{-i \varepsilon_{\mu}(\boldsymbol{k}) t} |u_{\mu}(\boldsymbol{k}) \rangle\,,\label{eq_state}
\end{equation}
where $c_\mu(\boldsymbol{k})=\langle u_\mu(\boldsymbol{k})|\xi_i\rangle$. Substituting Eq. (\ref{eq_state}) into Eq. (\ref{eq_texture}), we have
\begin{equation}
\begin{aligned}
\bar{\boldsymbol{n}}&=\lim _{T \rightarrow \infty} \frac{1}{T} \int_{0}^{T} \frac{\langle\xi(\boldsymbol{k})| \boldsymbol{\sigma} |\xi(\boldsymbol{k}) \rangle}{\langle\xi(\boldsymbol{k})|\xi(\boldsymbol{k}) \rangle} ~d t\\
&=\lim _{T \rightarrow \infty} \frac{1}{T} \int_{0}^{T} \frac{\sum_{\mu, \mu^{\prime}} c_{\mu} c_{\mu^{\prime}}^{*} e^{-i (\varepsilon_{\mu}-\varepsilon_{\mu^{\prime}} ) t} \langle u^R_{\mu^{\prime}} |\boldsymbol{\sigma} | u_{\mu}^R \rangle}{\sum_{\mu} |c_{\mu} |^{2} } d t .
\end{aligned}
\end{equation}
The off-diagonal sectors $\mu\neq \mu'$ are periodically oscillating functions and thus do not contribute to the long-time average \cite{BZhu2020,LZhou2019}. Then we have
\begin{equation}
\begin{aligned}
\bar{\boldsymbol{n}}_{x,y} &\cong \sum_{\mu} |c_{\mu} |^{2} \langle u^R_{\mu} |\sigma_{x,y} | u^R_{\mu} \rangle \\
&= ( |c_{+} |^{2}- |c_{-} |^{2} ) \frac{{\rm Re}(h_{x,y})\pm{\rm Im}(h_{y,x})}{\mathcal{N}_\mathbf{k}}
.\end{aligned}
\end{equation}
On the BIS, $|c_{+} |^{2}- |c_{-} |^{2}=0$, thus $\bar{\boldsymbol{n}}_{x,y}=0$. Then we calculate $\bar{\boldsymbol{n}}_z$. Calculation shows that
 \begin{equation}
\langle u^R_{-}|\sigma_z|u^R_{-}\rangle=\frac{4 d_z  (\varepsilon-d_z  )+2 a (a-b)  ( d_x^2+ d_y^2 )}{4 d_z  ( d_z-\varepsilon )+(a+b)^2  (d_x^2+d_y^2 )},
\end{equation}
\begin{equation}
\langle u^R_{+}|\sigma_z|u^R_{+}\rangle=\frac{-4 d_z  (\varepsilon+d_z  )+2 a (a-b)  ( d_x^2+ d_y^2 )}{4 d_z  ( d_z+\varepsilon )+(a+b)^2  (d_x^2+d_y^2 )}.
\end{equation}
For Hermitian case $a=b=1$, $\bar{\boldsymbol{n}}_z=0$ holds on the BIS since $\langle u^R_{-}|\sigma_z|u^R_{-}\rangle=-\langle u^R_{+}|\sigma_z|u^R_{+}\rangle$ and $|c_{+} |^{2}- |c_{-} |^{2}=0$. In contrast, when the non-Hermiticity is switched on, we have $\langle u^R_{-}|\sigma_z|u^R_{-}\rangle=\langle u^R_{+}|\sigma_z|u^R_{+}\rangle$ on the BIS where $d_z=0$, leading to $\bar{\boldsymbol{n}}_z=1$. This implies that only $\bar{\boldsymbol{n}}_{x,y}$ vanishes on the BIS for the pseudo-Hermitian case. When we quench the $d_{x,y}$ axis for a Hermitian Hamiltonian, the topological information can be restored with $\bar{\boldsymbol{n}}_{z}$ resting on the higher-order BIS \cite{WJia2021}. This method fails for the non-Hermitian quench dynamics due to the anomalous behavior of $\bar{\boldsymbol{n}}_{z}$.

The directional derivative of time-averaged Bloch vector $\bar{\boldsymbol{n}}$ is given by
\begin{equation}
\begin{aligned}
&\mathbf{g}(\mathbf{k}) \equiv-\frac{1}{\mathcal{N}_{\mathbf{k}}} \partial_{k_{\perp}} \bar{\boldsymbol{n}}=-\lim _{k_{\perp} \rightarrow 0^{+}} \frac{1}{\mathcal{N}_{\mathbf{k}}} \frac{\Delta \bar{\boldsymbol{n}}}{2 k_{\perp}}\\
&=-\lim _{k_{\perp} \rightarrow 0^{+}} \frac{1}{\mathcal{N}_{\mathbf{k}}} \frac{ \bar{\boldsymbol{n}} |_{d_{z} (k_{\perp} )>0}- \bar{\boldsymbol{n}} |_{d_{z} (-k_{\perp} )<0}}{2 k_{\perp}}.
\end{aligned}
\end{equation}
Near the BIS, $|c_{+} |^{2}- |c_{-} |^{2} \cong k_{\perp}$, then
\begin{equation}
\begin{aligned}
 &\Delta \bar{\boldsymbol{n}}_{x,y} |_{k_{\perp} \rightarrow 0}=- \{ \frac{[{\rm Re}~h_{x,y}(k_{\perp}, \boldsymbol{k}_{\|} ) \pm{\rm Im}~h_{y,x}(k_{\perp}, \boldsymbol{k}_{\|} )]  k_{\perp}}{k_{\perp}^{2}+\sum_{\mu=x,y} h^2_{\mu} (k_{\perp}, \boldsymbol{k}_{\|} )} \\
& -\frac{[{\rm Re}~h_{x,y}(-k_{\perp}, \boldsymbol{k}_{\|} ) \pm{\rm Im}~h_{y,x}(-k_{\perp}, \boldsymbol{k}_{\|} )] (-k_{\perp} )}{k_{\perp}^{2}+\sum_{\mu=x,y} h^2_{\mu} (k_{\perp}, \boldsymbol{k}_{\|} )} \} \\
&= -2 \frac{[{\rm Re}~h_{x,y}(0, \boldsymbol{k}_{\|} ) \pm{\rm Im}~h_{y,x}(0, \boldsymbol{k}_{\|} )] k_{\perp}}{\sum_{\mu=x,y} h^2_{\mu} (0, \boldsymbol{k}_{\|} )}+\mathcal{O} (k_{\perp}^{2} )\,.
\end{aligned}
\end{equation}
Now we obtain that
\begin{equation}
g_{x,y}(\boldsymbol{k})|_{\rm{BIS}}={\rm Re}(\hat{h}_{x,y})\pm{\rm Im}(\hat{h}_{y,x})=\hat{d}_{x,y}\,.
\end{equation}

\section{Floquet Hamiltonian}\label{appd}
For the square-wave driving, the effective Hamiltonian is given by $\mathcal{H}_{F}=i \log \left[e^{-i \mathcal{H}_{2}(\mathbf{k}) T_{2}} e^{-i \mathcal{H}_{1}(\mathbf{k}) T_{1}}\right] / T$ with $T_2=T-T_1$. We decompose the effective Hamiltonian in terms of the Pauli matrices (rather than the q-deformed matrices) $\mathcal{H}_{F}=\mathbf{d}_{\rm{eff}}(\mathbf{k})\cdot \boldsymbol{\sigma}$, with $\mathbf{d}_{\mathrm{eff}}(\mathbf{k})=-\arccos (\epsilon) \underline{\mathbf{r}} / T$ \cite{HWu2020},
\begin{equation}
\begin{aligned}
\varepsilon=&\cos  (T_{1} \varepsilon_{1}) \cos  (T_{2} \varepsilon_{2} )-\underline{\mathbf{d}}_{1} \cdot \underline{\mathbf{d}}_{2} \sin  (T_{1} \varepsilon_{1} ) \sin (T_{2} \varepsilon_{2} ),\\
\mathbf{r}=& \underline{\mathbf{d}}_{1} \times \underline{\mathbf{d}}_{2} \sin \left(T_{1} \varepsilon_{1}\right) \sin \left(T_{2} \varepsilon_{2}\right)-\underline{\mathbf{d}}_{2} \cos \left(T_{1} \varepsilon_{1}\right) \\
& \times \sin \left(T_{2} \varepsilon_{2}\right)-\underline{\mathbf{d}}_{1} \cos \left(T_{2} \varepsilon_{2}\right) \sin \left(T_{1} \varepsilon_{1}\right),
\end{aligned}
\end{equation}
where $\underline{\mathbf{d}}_{j}=\mathbf{h}_j/\varepsilon_j$,  $\varepsilon_j=\sqrt{\mathbf{h}_{j}\cdot \mathbf{h}_{j}}$, and $\mathbf{h}_{j}=(h_{j,x},h_{j,y},h_{j,z})$. The crossing of the decoupled Floquet subbands $r_z(\mathbf{k})=0$ [or crossing in all blocks $\mathcal{H}+2m\omega$ ($m=0,\pm 1,\pm 2,\ldots$)] defines the 0 BIS, while the crossings of the lower band in  $\mathcal{H}+2m\omega$ blocks and the upper band in $\mathcal{H}+(2m-1)\omega$ blocks define the $\pi$ BIS, i.e., $r_z(\mathbf{k})=\omega/2$.

\end{appendix}

\begin{appendix}

\end{appendix}

\end{document}